# Critical-like behavior of ionic-related, low-frequency dielectric properties in compressed liquid crystalline 8OCB and its nanocolloid

J. Łoś, A. Drozd-Rzoska, S. J. Rzoska

Institute of High Pressure Physics, Polish Academy of Sciences,

ul. Sokołowska 29/37, 01-142 Warsaw, Poland

Joanna Łoś:

ORCID: 0000-0001-6283-0948 ; e-mail: joalos@unipress.waw.pl

Aleksandra Drozd-Rzoska:

ORCID: 0000-0001-8510-2388 ; e-mail: Ola.DrozdRzoska@gmail.com

Sylwester J. Rzoska:

ORCID: 0000-0002-2736-2891 ; e-mail: sylwester.rzoska@gmail.com

**Abstract**

The report presents pressure-related broadband dielectric spectroscopy (BDS) studies in liquid crystalline octyloxycyanobiphenyl (8OCB) and its nanocolloid with $BaTiO_3$ nanoparticles, focused on the ionic-related low-frequency (LF) domain and the influence of pretransitional fluctuations. Hence, basic exogenic (pressure) and endogenic (nanoparticles) impacts on the dielectric properties are addressed. The innovative derivative-based analysis revealed functional 'critical-like' descriptions of ionic- contributions to dielectric permittivity and electric conductivity. The supplementary dielectric constant scan, yielding insight into the preferable dipole-dipole arrangements, is also presented.

Studies cover the puzzling case of complex liquid (Smectic A) – Solid crystal phase transition, revealing relatively strong, critical-like, premelting effects, which have hardly been observed for the discontinuous 'melting transition' so far.

**Introduction**

Soft matter constitutes the category of systems linking the dominance of mesoscale elements, often multi-molecular species, and the enormous sensitivity to perturbations [1-3]. In this family, rod-like liquid crystalline (LC) compounds play the reference model role [2, 4]. It is associated with the simple molecular 'geometry', well approximated by rods or ellipsoids, and the significance of pretransitional, multi-molecular, fluctuations [4-7]. Their size (correlation length, $\xi$) and lifetime $\left(\tau_{fl}\right)$ exhibit a critical-type growth on approaching mesophasic phase transitions [7]:

$$\xi(T) = \xi_0 |T - T^*|^{-\nu} \quad , \qquad \tau_{fl}(T) = \tau_0 |T - T^*|^{-z\nu} \tag{1}$$

where $T^*$ denotes the (extrapolated) hypothetical continuous phase transition temperature, $\nu$ is the universal correlation length exponent, and $z$ stands for the scaling dynamical exponent. Values of critical exponents are described by the space and order parameter dimensionalities. The uniqueness of phase transition in LC systems is worth stressing: they are often related to single elements of symmetry freezing or releasing [7]. The continuous or weakly discontinuous character of phase transitions in LC systems leads to the significant influence of pretransitional fluctuations on material properties. It can be exemplified by dielectric constant changes [8-22]:

$$\varepsilon(T) = \varepsilon^* + a|T - T^*| + A|T - T^*|^{\phi = 1-\alpha} \qquad \text{for} \quad T = const \tag{2}$$

where $\alpha$ is the heat capacity 'critical' exponent; $\alpha = 1/2$.

Eq. (2) shows that pretransitional effects depend on the distance from the continuous phase transition temperature $(T^*)$ and are governed by universal critical exponents. For weakly-discontinuous phase transitions at $T = T_D$, the discontinuity metric $\Delta T^* = |T_D - T^*|$ is introduced. The appearance of pretransitional effects is possible if a 'contrast factor' between fluctuations and their surrounding exists. For instance, in the isotropic liquid phase of nematic LC materials dielectric constant within prenematic fluctuations can be significantly lesser than for the isotropic surrounding. It occurs for LC molecules with the permanent dipole moment parallel to the long molecular axis. In such a case, the uniaxial prenematic ordering within fluctuations leads to the accidental 'up' and 'down' positioning of permanent dipole moments and the cancellation of the dipolar contribution to dielectric constant, which becomes lesser than for the isotropic surrounding. Consequently, on approaching the Isotropic - Nematic (I-N) transition, the crossover $d\varepsilon(T)/dT < 0 \to d\varepsilon(T)/> 0$, described by Eq. (2), occurs. The contrast factor is absent for rod-like LC molecules with the permanent dipole moment approximately perpendicular to the long molecular axis. In such a case, $\varepsilon(T)$ pretransitional effect does not appear [12]. Notable that fluctuations-driven changes of dielectric constant in the isotropic liquid phase are significant even 100 K above I-N transition [14-17, 21, 22]. They also dominate the whole range of LC mesophases, as shown recently [20- 22].

The sensitivity of LC systems to perturbations can illustrate the strong impact of such canonic exogenic factors as pressure. Even moderate-range compressing notably changes phase transitions temperatures and $\Delta T^*(P)$ values [15, 19, 21]. Regarding pretransitional effects, their functional forms on cooling and compressing are isomorphic, in agreement with the *Physics of Critical Phenomena* [7]. Notable that pressure-related studies in LC-based nanocolloids are still very limited [23].

As for the endogenic impact factors, adding even a tiny amount of solid nanoparticles (NPs) can significantly change the LC host properties [24, 25]. Unique features of LC-based nanocolloids led to enormous interest in such materials and the creation of a new category in the *Physics of Liquid Crystals* [24, 25]. However, studies on the impact of pretransitional fluctuations on such nanocoloids are still limited [23, 26-32].

Changes in dielectric constant in LC systems illustrate the above discussion. Its is associated with broadband dielectric spectroscopy (BDS) [33] monitoring in the frequency range $1kHz < f < 1MHz$. The impact of fluctuations is also explicitly manifested in high-frequency BDS studies up to the GHz domain [14-17, 30-31].

Numerous research reports are focused on dielectric properties of LC compounds and their nancocolloids in the ionic-related low-frequency (LF) domain, i.e., for frequencies below the static domain [24, 25, 32, 34-56]. It is motivated by omnipotent applications of liquid crystalline materials, for which the enormous sensitivity to the external electric field and LF dielectric properties are essential [4, 6, 57]. For nanocolloids this interest is associated with new generations of displays and photonic devices [23, 24, 57].

However, the insight into the impact of pretransitional fluctuations on dielectric properties detected in the low-frequency BDS studies remains poor. It is associated with the practical lack of temperature or pressure evolutions of reference dielectric properties in this domain. Such evolutions were considered only in the isotropic phase of pentylcyanobiphenyl (5CB) [13] and, very recently in undecylcyanobiphenyl (11CB) and its nanocolloids [32]. In 11CB the influence of fluctuations was also detected in SmA mesophase, although only qualitatively.

This report presents pressure-related broadband dielectric spectroscopy (BDS) studies in liquid crystalline octyloxycyanobiphenyl (8OCB) and its nanocolloid with $BaTiO_3$ nanoparticles, focused on the ionic-related low-frequency (LF) domain and the influence of pretransitional fluctuations. The innovative derivative-based analysis revealed the critical-like form of the ionic-related contribution to dielectric permittivity and electric conductivity. Studies cover Nematic, Smectic A mesophases and extend to the liquid SmA–Solid crystal transition, introducing new results to the challenging field of the discontinuous liquid-solid 'melting' transition and premelting effects [58-66].

**Experimental**

Studies were carried out in 4'-octyloxy-4-cyanobiphenyl (8OCB, octyloxycyanobiphenyl), the rod-like LC compound with $Crystal\ (Cr) \leftarrow 323.3K \leftarrow SmecticA\ (SmA) \leftarrow 340.4\ K \leftarrow Nematic\ (N) \leftarrow 353.6\ K \leftarrow Isotropic\ (I)$ mesomorphism [4, 31]. Its molecular structures is rod-like type, with the permanent dipole moment approximately parallel to the central part of the molecule. Notable that the alkyl chain 'tail' is slightly angled (72º) to the central part based on two phenyl rings, as shown schematically in the inset in Fig. 3. The effective dipole moment obtained from dielectric constant measurements and numerical modeling: $\mu = 7.3D$ [4, 67]. 8OCB samples were deeply purified to reduce the redundant ionic contaminations and degassed immediately before experiments.

Barium titanate paraelectric (cubic phase), globular $BaTiO_3$ nanopowder (diameter $2r = 50$ nm) was purchased from US Research Nanomaterials Inc. Nanocolloids (with a mass fraction of NPs) [68] were obtained due to one-hour sonication in the isotropic liquid phase. After this process, no sedimentation was observed in microscopic tests.

Samples were placed in a flat-parallel capacitor (diameter $2r = 16$ mm) made of the gold-coated Invar with the distance between plates $d = 0.15$ mm. The voltage of the measuring field $U = 1$ V was applied. BDS studies were carried out via Novocontrol Alpha-A analyzer in the frequency range from single Hz up to ca. 10 MHz. The latter determines the terminal value available in high-pressure studies. The high-pressure setup, with a computer-controlled pressure source, was designed and built by Unipress Equipment. In experiments, attention was paid to avoiding contamination from the liquid transmitting pressure changes (Plexol), since even the slightest impurity can significantly change the properties of tested systems. The pressure chamber was surrounded by a special 'jacket' linked to Julabo large-volume thermostat with external circulation for temperature control. The temperature stabilization was better than $\pm 0.02 K$, and monitored by the copper-constantan thermocouple within the pressure chamber and two Pt100 resistors placed in the body of the pressure chamber. Examples of obtained spectra presented via the real and imaginary part of the complex dielectric permittivity are shown in Fig. 1. Characteristic domains are indicated. The static domain is associated with the horizontal region of $\varepsilon'(f)$, where the frequency shift has a negligible impact on detected values. In this domain, dielectric constant is determined, i.e., $\varepsilon'(f) = \varepsilon$. The ionic-related low-frequency (LF) domain extends for frequencies below the static domain. It is dominated by the growing impact of the ionic polarizability associated with transport processes. For frequencies above the static domain, the impact of the primary ($\alpha$-) relaxation process emerges, manifested via $\varepsilon''(f)$ primary relaxation loss curve. Its peak determines the primary ($\alpha$-) relaxation time $\tau = 1/2f_{peak}$ [33]. The test of these properties for high-pressure studies presented in this report is not possible because of the mentioned frequency range cutoff.

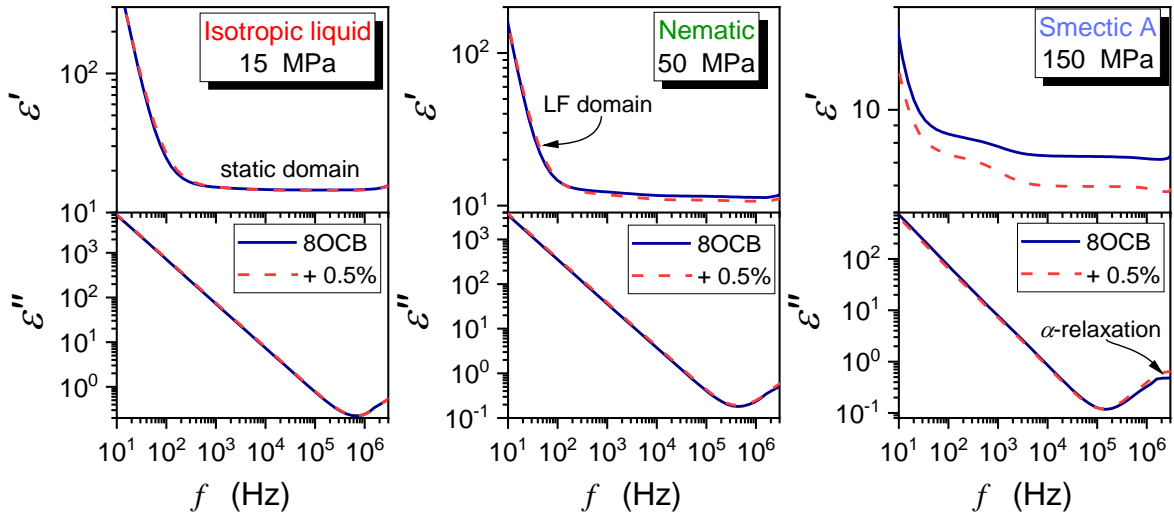

**Figure 1**    Examples of frequency spectra for the real and imaginary parts of dielectric permittivity in 8OCB and its nanocolloid with BaTiO$_3$ nanoparticles.

Figure 2 presents the part of data from Fig.1, transformed into the real part of the electric conductivity representation: $\sigma'(f) = 2\pi f \varepsilon''(f)$ [30]. The horizontal region in Fig. 2 is related to DC electric conductivity. The decrease in values below the DC reference, for $f < 10 Hz$, can be linked to the accumulation of electric charges in the immediate vicinity of the capacitor plates. Notable are disturbances in the DC-domain, revealed in the inset in Fig. 2. The influence of nanoparticles is also visible. One of 'classic' routines used in determining DC electric conductivity is the application of the Nyquist plot, yielding the zero-frequency extrapolation [33]. Its application was essential when studies were carried out using different experimental facilities, which could introduce bias errors. However, it does not enable detailed insights into subtle processes shown in Figs. 2 and 3, based on BDS scans possible nowadays. DC electric conductivity is also often determined using simply $\varepsilon''(f)$ the plot presented in Fig.1, via the dependence: $log_{10}\varepsilon''(f) = a + blog_{10}\omega$, where $\omega = 2\pi f$ and $a = \sigma'(f)$. The DC electric conductivity is related to $b = -1$ [33].

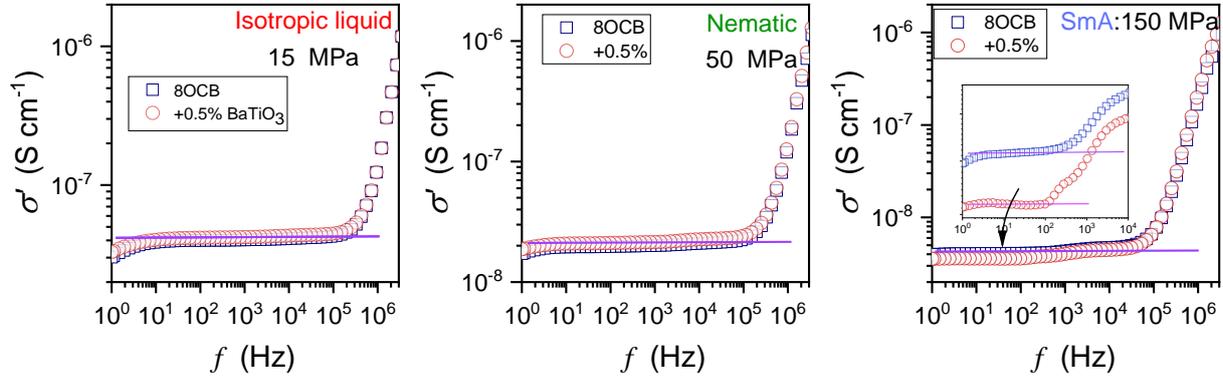

**Figure 2** Spectra of the real part of electric conductivity in tested 8OCB and its nanocolloid.

Rod-like LC compounds are often studied in samples oriented by a strong external magnetic for detecting properties associated with the parallel and perpendicular molecular axis orientations [4, 6]. Such an approach is not possible for high-pressure studies. Another experimental practice is the orientation of LC molecules by covering capacitor plates with the polymeric agent supporting the desired molecular orientation. Such studies require supplemental modeling to minimize the impact of the additional polymeric layer and have to be carried out in thin, micrometric layers [4, 6]. This report focuses on bulk material properties. It is also important, that pretransitional fluctuations can be disturbed by the thin-layer geometrical restrictions [6, 67]. Notably, phase transitions and pretransitional effects in LC materials are extremely sensitive even to tiny contaminations [4, 69, 70]. The polymeric agent covering capacitor plates can introduce such risk. The solid phase of LC materials and their nanocolloids are still poorly known. Figure 3 shows the emergence of a relaxation process in this domain due to the addition of nanoparticles.

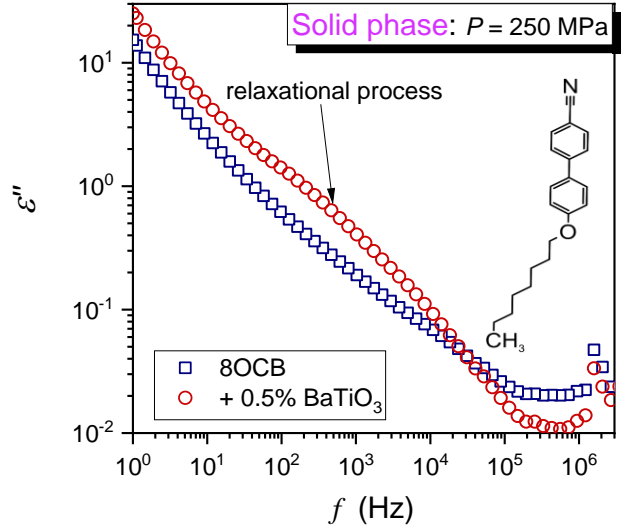

**Figure 3**  Frequency spectra of the imaginary part of dielectric permittivity in the solid phase of 8OCB.

The authors stress that the impact of discussed above 'disturbations' were taken into account in determining the DC electric conductivity. For $\varepsilon'(f)$ spectra the static domain was identified for each tested pressure, and dielectric constant was determined for the mid of this domain. The common practice of estimating dielectric constant for a single, selected frequency was avoided. In the given case, it could bias results.

**Results and Discussion**

Dielectric constant is the basic dielectric characterization of materials, which was introduced yet by Michel Faraday in the mid of 19[th] century [71, 72]. It reflects molecular polarizability and intermolecular interactions, particularly in systems with permanent dipole moments. In the low-frequency domain, the real part of dielectric permittivity is given as $\varepsilon'(f) = \Delta\varepsilon'_{ion} + \varepsilon_{dip} + \varepsilon_\infty$ [73]. When decreasing frequency, the component associated with the translation of ionic species becomes more and more important, and finally $\Delta\varepsilon'_{ion} \gg \varepsilon + \varepsilon_\infty$. When increasing frequency above the impact of $\Delta\varepsilon'_{ion}$ diminishes, and finally, in the static domain $\varepsilon'(f) = \varepsilon = \varepsilon_{dip} + \varepsilon_\infty$, where $\varepsilon_{dip}$ is related to permanent dipole moments, and $\varepsilon_\infty$ is for the high-frequency contribution due to electronic and vibrational processes. On further increasing frequency above the static domain, the impact of $\varepsilon_{dip}$ diminishes and then $\varepsilon'(f) \approx \varepsilon_\infty$, with typical values $\varepsilon_\infty \approx 2-3$. Classic models of *Dielectric Physics* focus on the phenomena associated with dielectric constant, and do not enter the LF domain. The latter strongly depends on the tested sample purity, which is particularly important in liquid dielectrics. 'Classic' dielectric properties are most often analyzed within Kirkwood and Froelich local field models, yielding a relation convenient for commenting experimental data [73-75]:

$$\frac{9k_B T}{4\pi\rho} \frac{(\varepsilon-\varepsilon_\infty)(2\varepsilon+\varepsilon_\infty)}{\varepsilon(\varepsilon_\infty+2)^2} = g_K \mu^2 \qquad (3)$$

where $\rho$ is for density of dipolar components; the Kirkwood factor $g_K < 1$ indicates a preferred antiparallel correlation between permanent dipole moments and $g_K > 1$ implies preferred parallel permanent dipole moment ($\mu$) correlations.

The above relation does not allow describing temperature changes of dielectric constant, but it is possible for Kirkwood factor, tested in LC materials, including 8OCB [76-80]. However, Kirkwood and Froelich local field models do not consider critical collective processes [73-75]. Notwithstanding the analysis of Eq. (3), supported by the experimental results, shows that the knowledge of temperature changes of dielectric constant offers quantitative insight into the dominant arrangements of permanent dipole moments. Namely, the evolution described by $d\varepsilon(T)/dT > 0$ suggests the preference for the 'antiparallel' dipole-dipole arrangement, and $d\varepsilon(T)/dT < 0$ suggests the 'parallel' one [73]. The latter means that individual permanent dipole moments follow the direction of the electric field.

Figure 4 shows pressure changes of dielectric constant in 8OCB and its BaTiO$_3$ based nanocolloid. Studies were conducted $\Delta T = 5K$ above the clearing (I-N transition) temperature. Under such conditions, the I-N transition occurs when compressing by $P = 20 MPa$ for 8OCB and $P = 25 MPa$ in the nanocolloid. Worth recalling that in temperature tests under atmospheric pressure, even adding 1% of the same NPs causes a negligible shift in the clearing temperature.

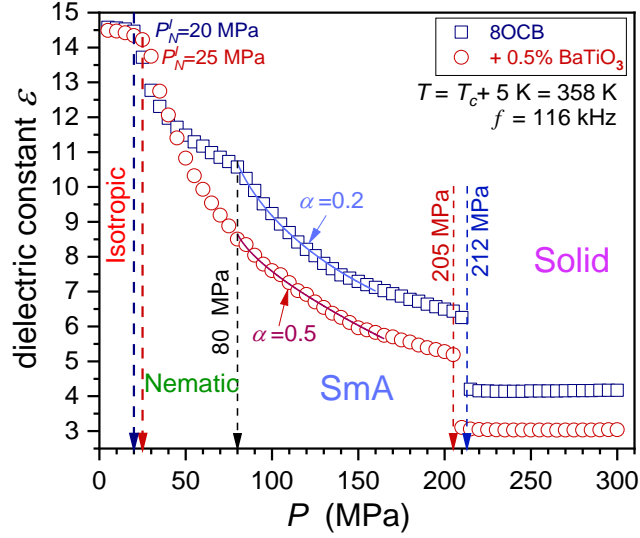

**Figure 4** The pressure evolution of dielectric constant in 8OCB and its nanocolloid, covering isotropic liquid, LC mesophases, and the solid phase.

One can consider the following pressure-related parallel of Eq. (2):

$$\varepsilon(P) = \varepsilon^* + a|P^* - P| + A|P^* - P|^{\phi=1-\alpha} \qquad \text{for} \qquad P = const \qquad (4)$$

The authors recall the isomorphism postulate for critical phenomena, indicating parallel forms of precritical effect for temperature and pressure approaching the critical point [7]. For results presented in Fig. 4, Eqs. (4) cannot portray $\varepsilon(P)$ in the isotropic liquid phase since it is available only in a limited range of pressures. Notwithstanding the region of increasing antiparallel arrangement on approaching $P_N^I$ is visible, particularly in Figure 5, presenting the derivative of experimental data from Fig. 4. In the nematic phase, the gradual decrease of dielectric constant, suggests that compressing favors the arrangement of a long molecular axis perpendicular to the direction of the electric field, i.e., parallel to capacitor plates. This process seems to be more effective in the nanocolloid, as visible in Fig. 4. In Smectic A phase, the sensitivity to uniaxial ordering by external factors diminishes, and the analysis of pretransitional effect is possible. The application Eq. (2) yielded the exponent $\alpha \approx 0.2$ for pure 8OCB and $\alpha \approx 0.5$ for the nanocolloid, with the extrapolated singular pressure $P^* \approx 77\ MPa$. These values correlate with the evidence for the $N \leftarrow SmA$ transition, showing the similar span between values characteristic for the XY Ising universality class and mean field for the critical phenomena physics [7]. The fitting results, validating the discussed parameterization, are given in Fig. 4.

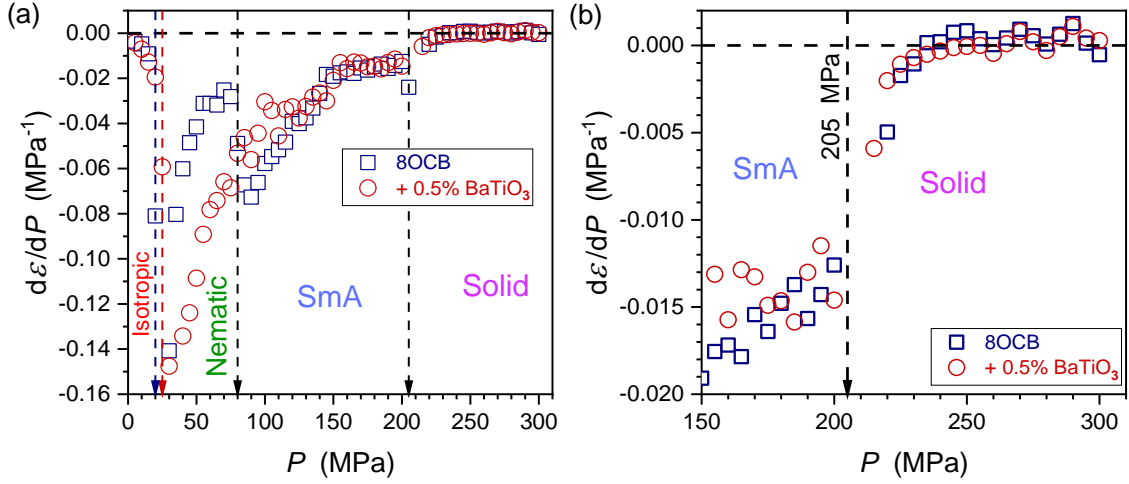

**Figure 5** The derivative analysis of the dielectric constant pressure evolution data given in Fig. 4, for 8OCB and its nanocolloid. The left part is for the tested pressure range, and the right one is focused on better insight into the ordered fluid (SmA) – Solid phase transition.

Figure 4 and particularly Figure 5 show that neither nematic nor Smectic A phase are uniform: they split into domains associated with vicinities of subsequent phase transitions. Interestingly, the derivative of dielectric constant reveals hallmarks of the pretransitional effect also for the $SmA - S$ transition. In the solid phase, it is associated with an increasing tendency for the antiparallel arrangement of permanent dipole moments, as shown by $d\varepsilon(P)/dP$ evolution tendency. Figure 6 presents pressure changes of the real part of dielectric permittivity for a set of frequencies, from the static reference to the low-frequency (LF), ionic-related domain. The lowering of the monitoring frequency leads to the essential increase of $\varepsilon'(f)$ values. For the analysis of emerging phenomena, one can focus on the ionic contribution, namely:

$$\Delta\varepsilon'(f,T,P) = [\varepsilon'(f,T,P)]_{LF} - [\varepsilon'(f,T,P)]_{Static} \qquad (5)$$

Such analysis carried out for the temperature evolution, for $P = 0.1 MPa$, in the isotropic phase of pentylcyanobiphenyl (5CB, I-N transition) [13] and undecylcyanobiphenyl (11CB, I-SmA transition) [32] revealed linear changes of $\Delta\varepsilon'(f,T)$.

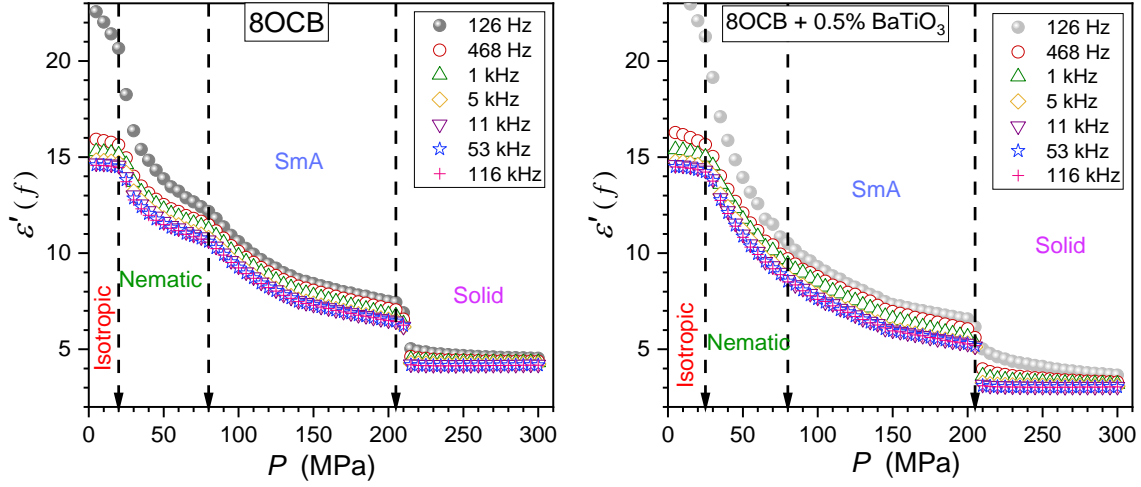

**Figure 6**    The pressure evolution of the real part of dielectric constant in 8OCB (left part) and its nanocolloid (right part) for frequencies ranging from the static domain down to the low frequency (LF) region.

This pattern was linked to the structured impact of premesomorphic fluctuations. In 11CB the impact of pretranstional fluctuations was also detected in SmA mesophase, but solely visually. This report presents the first attempt to test the possible functional form of the ionic contribution to dielectric permittivity based on the data given in Fig. 6. The authors assumed the critical-like form of such changes:

$$\Delta\varepsilon'(f,P) = A|P - P^+|^{-\varphi} \qquad (6)$$

where $T = const$, $A$ is the constant amplitude, and $P^+$ is a singular temperature for transport-related ionic processes.

The above relation splits into two equations, depending on whether the considered phase transition is reached on compressing or decompressing. For $P \to P^+$, as for $I \to N$, $N \to SmA$, and $SmA \to S$ pretransitional effects, one obtains:

$$\Delta\varepsilon'(f,P) = A(P^+ - P)^{-\varphi} \quad \Rightarrow \quad ln\Delta\varepsilon'(f,P) = lnA - \varphi ln(P^+ - P) \quad \Rightarrow$$

$$\Rightarrow \quad \frac{dln\Delta\varepsilon'(f,P)}{dP} = \frac{-\varphi}{P^+ - P} \quad \Rightarrow \quad \left[\frac{dln\Delta\varepsilon'(f,P)}{dP}\right]^{-1} = -\frac{P^+}{\varphi} + \varphi^{-1}P = \pm a \mp bP \quad (7)$$

where $P^+$ is the extrapolated singular pressure determined from the condition $[dln\Rightarrow\varepsilon'/dP]^{-1} = 0$, and the exponent $\varphi = 1/b$.

For $P^+ \leftarrow P$, as in the case of $I \leftarrow N$, $N \leftarrow SmA$, and $SmA \leftarrow S$ pretransitional effects:

$$\Delta\varepsilon'(f,P) = A(P - P^+)^{-\varphi} \quad \Rightarrow \quad ln\Delta\varepsilon'(f,P) = lnA - \varphi ln(P - P^+) \quad \Rightarrow$$

$$\Rightarrow \quad \frac{dln\Delta\varepsilon'(f,P)}{dP} = \frac{-\varphi}{P - P^+} \quad \Rightarrow \quad \left[\frac{dln\Delta\varepsilon'(f,P)}{dP}\right]^{-1} = \frac{P^+}{\varphi} - \varphi^{-1}P = a - bP \quad (8)$$

where exponent $\varphi = 1/b$.

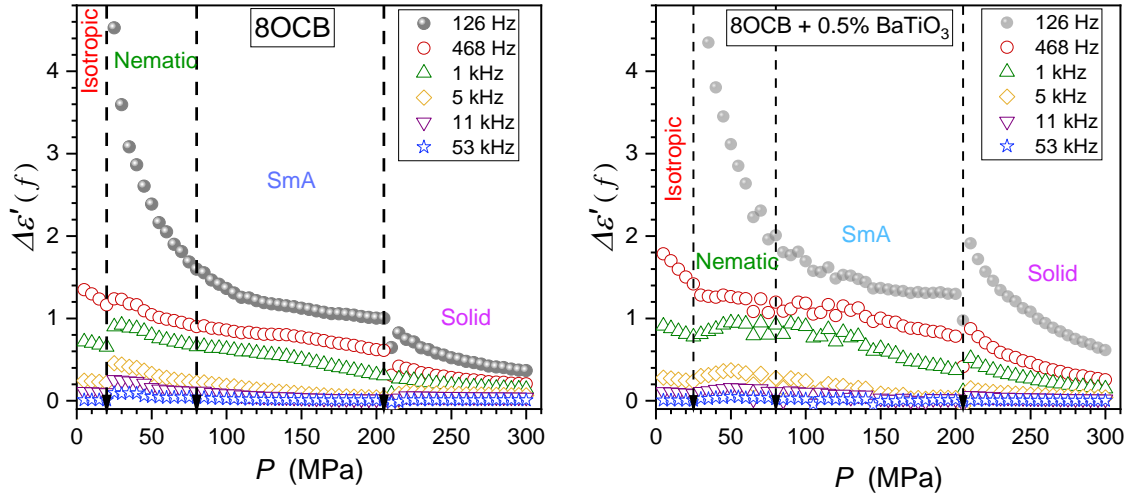

**Figure 7** The presentation of emerging LF contribution to the real part of dielectric permittivity for 8OCB (left part) and its nanocolloid (right part).

Figures 8 and 9 present $\Delta\varepsilon'(f,P)$ experimental data from Figs. 7a and 7b transformed according to Eqs. (7) and (8). The set of explicit linear domains in subsequent phases of tested 8OCB and its nanocolloids constitute a superior confirmation of the prevalence for the portrayal of the ionic contribution via the critical-like Eq. (6). Values of exponents and singular temperatures are given in Fig.8 (8OCB) and Fig. 9 (8OCB+NPs).

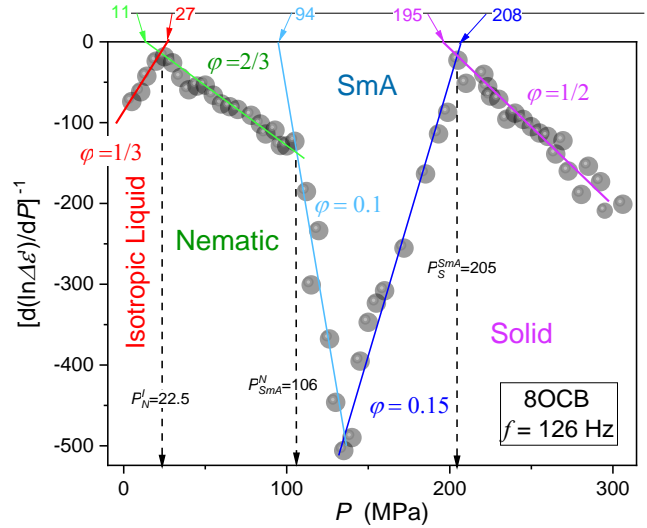

**Figure 8**  The derivative-based analysis (Eqs. (7) and (8)) focused on validating the 'critical-like' description via Eq. (6) of the LF ionic contribution to the dielectric permittivity in subsequent phases of 8OCB. Linear domains validate a preference for critical-like portrayal. Values of exponents and singular temperatures are given in the plot.

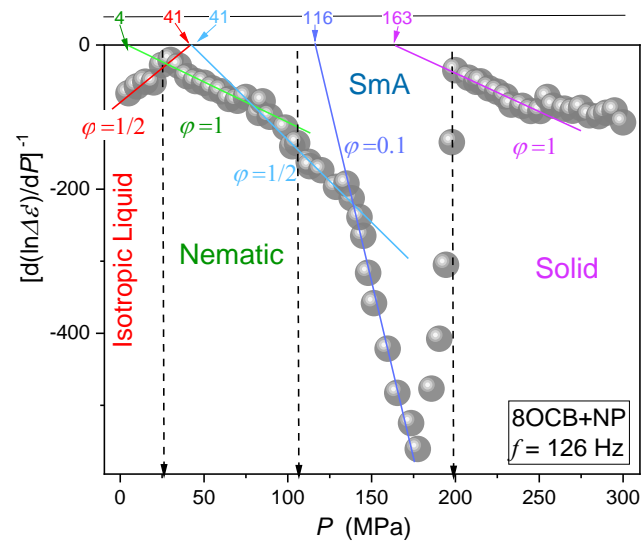

**Figure 9**  The derivative-based analysis (Eqs. (7) and (8)) focused on validating the 'critical-like' description via Eq. (6) of the LF ionic contribution to the dielectric permittivity in subsequent phases of 8OCB+BaTiO$_3$ nanocolloid. Linear domains validate the preference for critical-like portrayal. Values of exponents and singular temperatures are given in the plot.

DC electric conductivity constitutes another dielectric property characterizing the LF domain [33], in the given case, associated with the imaginary part of dielectric permittivity, as discussed above. Generally, in soft matter systems, the decay $\sigma(T,P)$ on cooling or compressing is expected. The best examples are ultraviscous glass-forming systems on approaching the glass temperature $(T_g)$ or pressure $(P_g)$. For temperature changes, such behavior is considered in frames of the Arrhenius (A) or Super-Arrhenius (SA) relation [33]. The latter, often observed in complex liquids and soft matter systems, is governed by the temperature-dependent apparent activation energy $E_a(T)$: $\sigma(T) = \sigma_\infty exp(-E_a(T)/RT)$, where $P = const$ and $R$ stands for the gas constant [33].

For isothermal compressing, the parallel Super-Barus behavior is expected [81-83]:

$$\sigma(P) = \sigma_0 exp\left(P\frac{-V_a(P)}{RT}\right) \quad \Rightarrow \quad \sigma^{-1}(P) = \rho(P) = \rho_\infty exp\left(P\frac{V_a(P)}{RT}\right) \quad (9)$$

where $T = const$, $V_a(P)$ is the apparent activation volume. The basic Barus pattern is retrieved for $V_a(P) = V_a = const$ in the given pressure domain. In such a case, a linear behavior appears in the plot defined by Eq. (9).

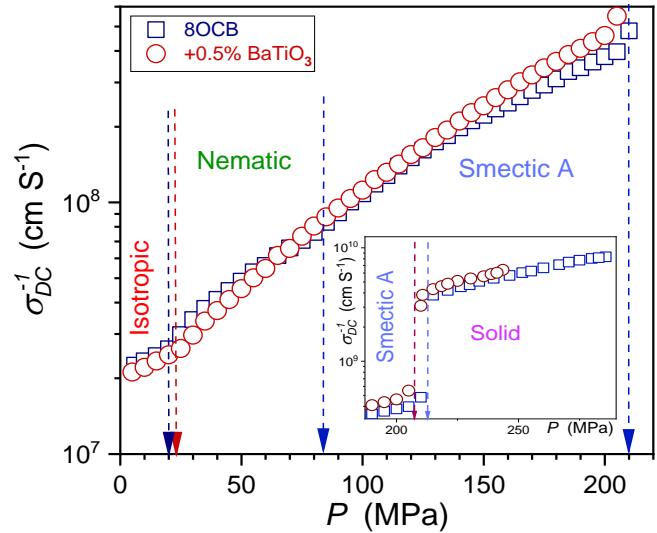

**Figure 10** The pressure evolution of DC electric conductivity in LC mesophases of 8OCB and its BatiO$_3$ related nanocolloid. The inset is focused on the solid-state behavior.

In soft matter systems, changes in dynamics properties on cooling or compressing are most often illustrated by viscosity or the primary relaxation time evolutions. Both properties show a parallel pattern of the SA/SB slowing down in relaxation time or increasing viscosity. Similar behavior is obtained when analyzing the reciprocal of electric conductivity $\sigma^{-1}(T,P)$, i.e., the electric resistivity [33, 81]. Notable that typical changes of $\tau(T,P)$, $\eta(T,P)$ or $\sigma^{-1}(T,P)$, for instance, in glass-forming systems, are always above the reference the basic reference ($E_a = const$) or Barus ($V_a = const$) patterns for subsequent temperatures on cooling or pressures on compressing, respectively [33, 81-83].

The evolutions of the electric conductivity reciprocal in 8OCB and the related nanocolloid are shown in Figure 11. Reported changes are below the reference Barus pattern, i.,e. for subsequent pressures on compressing the steepness $V^{\#}(P) = dln\sigma^{-1}(P)/dP$ decreases. To get insight into the hidden features of results presented in Fig. 10, the authors would like to recall the results of ref.[ ], where the 'universal' evolution of the so-called pressure fragility $m_T(P)$ on compressing in slowing down, glass-forming liquids were reported [82, 83]:

$$m_T(P) = \frac{M}{P^* - P} \qquad (10)$$

where $P^*$ is the extrapolated singular pressure and $M$ is the constant amplitude.

Taking into account the definition of apparent fragility, one can directly link it to the steepness index, namely [82, 83]:

$$m_T(P) = \frac{dlog_{10}\tau(P)}{d(P/P_g)} = \frac{P_g}{ln10}\frac{dln(P)}{dP} = cV^{\#}(P) \qquad (11)$$

where $c = const$

In ref. [82], by linking the definition of the apparent fragility (Eq. (11)) and the empirical finding given by Eq. (11) the critical-like relation for the relaxation time was derived:

$$\tau(P) = \tau_0(P - P^*)^{-\Gamma} \qquad (12)$$

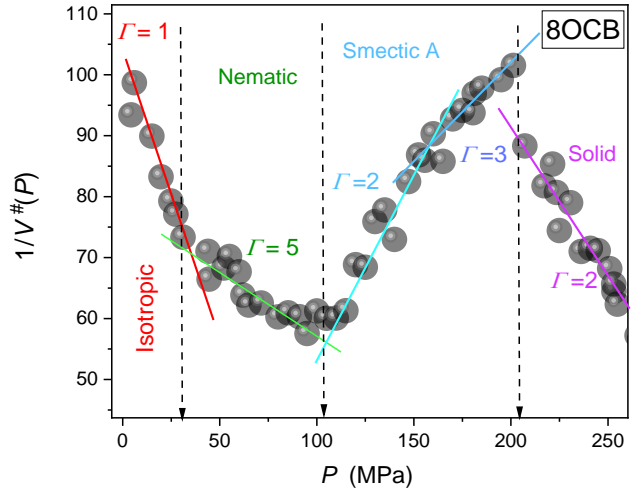

**Figure 11**   The reciprocal of the steepness index $V^{\#}(P) = dln\sigma^{-1}(P)/dP$ for electric conductivity changes given in Fig. 10, for 8OCB. Note the link to Eqs. (13) and (14).

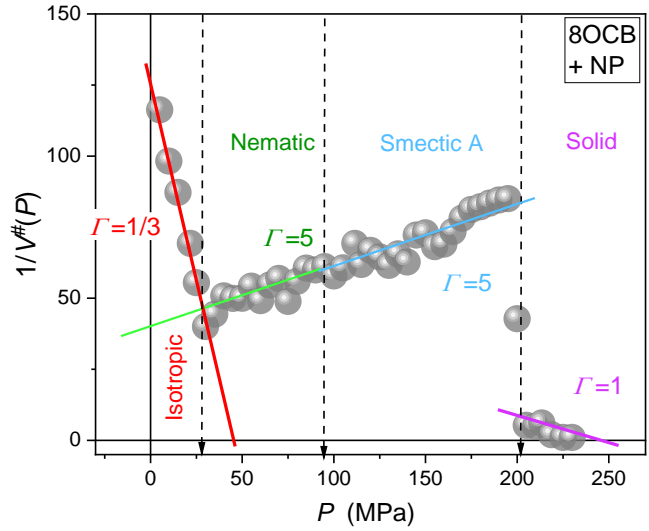

**Figure 12**   The reciprocal of the steepness index $V^{\#}(P) = dln\sigma^{-1}(P)/dP$ for electric conductivity changes given in Fig. 10, for 8OCB+ BaTiO$_3$ nanocolloid. Note the link to Eqs. (13) and (14).

Applying the above to changes in electric conductivity, one obtains:

$$\sigma^{-1}(P) = \rho_0|P - P^+|^{-\Gamma} \tag{13}$$

where $P^+$ is the extrapolated singular pressure.

The above relation is coupled to the following distortions sensitive test:

$$V^{\#}(P) = \frac{d\ln\sigma^{-1}(P)}{dP} = \frac{V}{|P-P^+|} \qquad (14)$$

where $V$ is the constant amplitude.

It leads to two linear checkpoint equations, depending on whether the phase transition is reached by compressing or decompressing:

$$[V^{\#}(P)]^{-1} = \frac{P^+}{\Gamma} - \Gamma^{-1}P = a - bP \qquad (15)$$

$$[V^{\#}(P)]^{-1} = -\frac{P^+}{\Gamma} + \Gamma^{-1}P = -a + bP \qquad (16)$$

Figures 11 and 12 show superior linear evolutions of $[V^{\#}(P)]^{-1}$, as suggested by Eqs. (15) and (16), what explicitly validates the critical-like portrayal of $\sigma^{-1}(P)$ by Eq. (14) in all phases of 8OCB and 8OCB+NPs nanocolloid.

The magnitude $V^{\#}(P)$, determined from pressure evolutions of different dynamical properties, is often considered the equivalent of the apparent activation volume appearing in the Super-Barus equation. However, the SB equation directly yields [82, 83]:

$$RT\, d\ln\sigma^{-1}(P)/dP = RTV^{\#}(P) = V_a + P\frac{dV_a(P)}{dP} \qquad (17)$$

Hence, generally $V^{\#}(P) \neq V_a(P)$. They coincide only for $P = 0$ or for the basic Barus behavior with $V_a(P)=V_a$ in the given pressure domain.

**Conclusions**

The report reveals the evolution of dielectric properties in the ionic-related, low-frequency domain of liquid crystalline compound (8OCB) and its nanocolloid, which is lacking so far. Its critical-like forms are in the isotropic liquid, nematic, smectic A, and solid phase, indicating the possible dominant influence of pretransitional fluctuations. Significant was the introduction of the innovative, distortions-sensitive analysis. The analysis was carried out for ionic-related contributions to the real part of dielectric permittivity $\Delta\varepsilon'(f)$, and associated with the imaginary part DC electric conductivity $\sigma = (2\pi f)\varepsilon''(f)$, for frequencies below the static domain. Results presented can also be considered as the first coherent test of the exogenic (high pressure) and endogenic (nanoparticles) impacts on LC compound, in the 'ionic' and the reference static domains. The supplementation of these results by the reference evidence for critical-like changes of dielectric constant are also presented.

The nature of the 'melting' $Liquid - Solid$ discontinuous phase transition remains a grand cognitive challenge [58-66]. Reference results show weak premelting effects, detected in temperature tests and impossible for functional analysis. Two exceptional cases of dielectric constant premelting changes in menthol [64] and thymol [65] (non-LC materials) have been reported very recently.

This report presents the first-ever pressure test of the (complex) liquid (SmA) – Solid phase transition, showing relatively strong critical-like premelting effects for $\Delta\varepsilon'(f)$ and $\sigma^{-1}(P)$ tests, assisted by $\varepsilon(P)$ changes. The results presented in this report indicate that the new experimental cognitive gate associated with ionic-related LF domain in BDS studies, poorly explored so far.

**Declaration of Competing Interest**

The authors declare that they have no known competing financial interests or personal relationships that could have appeared to influence the work reported in this paper.

**Authors Contributions**

JŁ was responsible for measurements, data analysis, and paper preparations, ADR supported data analysis and paper preparation, SJR supported paper preparation.

**Experimental data availability:**

The author declares the availability of experimental data upon reasonable request.

**Authors reference data:**

Joanna Łoś: ORCID: 0000-0001-6283-0948 ; e-mail: joalos@unipress.waw.pl

Aleksandra Drozd-Rzoska:

ORCID: 0000-0001-8510-2388 ; e-mail: Ola.DrozdRzoska@gmail.com

Sylwester J. Rzoska: ORCID: 0000-0002-2736-2891 ; e-mail: sylwester.rzoska@gmail.com

**References**

1. P.G. de Gennes, J. Badoz, Fragile Objects: Soft Matter, Hard Science, and the Thrill of Discovery, Copernicus, NY, 1996.

2. R.A.L. Jones, Soft Condensed Matter, Oxford Univ. Press, Oxford, 2002.

3. M. Doi, Soft Matter Physics, Oxford Univ. Press., Oxford, 2013.

4. D. Demus, J. Goodby, G.W. Gray, H.-W. Spiess, and V.Vill, Handbook of Liquid Crystals: Fundamentals,Wiley-VCH, Weinheim, 1998.

5. P. Pieranski, Pierre-Gilles de Gennes: beautiful and mysterious liquid crystals, Comp. Rend. Physique 20 (2019) 756-769.

6. P. Collings and J.W. Goodby, Introduction to Liquid Crystals: Chemistry and Physics, CRC Press, Routlege, 2019.

7. M.A. Anisimov, Critical Phenomena in Liquids and Liquid Crystals, Gordon and Breach, Reading, UK, 1992.

8. J. Thoen, G. Menu, Temperature dependence of the static relative permittivity of octylcyanobiphenyl (8CB), Mol. Cryst. Liq. Cryst. 97 (1983) 163–176.

9. A. Drozd-Rzoska, S.J. Rzoska, J. Zioło, Critical behavior of dielectric permittivity in the isotropic phase of nematogens. Phys. Rev. E 54 (1996) 6452-6456.

10. A. Drozd-Rzoska, S.J. Rzoska and J. Zioło, The influence of high pressure on the discontinuity of the isotropic - nematic and isotropic - Smectic A transitions studies by the low - frequency nonlinear dielectric effect measurements, Mol. Cryst. Liq. Cryst. 330 (1999) 7-14.

11. A. Drozd-Rzoska, S.J. Rzoska, J. Zioło, The fluid-like and critical behavior of the isotropic-nematic transition appearing in linear and non-linear dielectric studies. Acta Phys. Polon. A 98 (2000) 637–643.

12. A. Drozd-Rzoska, S. Pawlus, S.J. Rzoska, Pretransitional behavior of dielectric permittivity on approaching a clearing point in mixture of nematogens with antagonistic configurations of dipoles, Phys. Rev. E 64 (2001) 051701.

13. A. Drozd Rzoska, S.J. Rzoska, J. Zioło, J. Jadżyn, Quasi-ciritical behavior of the low-frequency dielectric permittivity in the isotropic phase of n-pentylcyanobiphenyl, Phys. Rev. E 63 (2001) 052701 (4).

14. S.J. Rzoska, M. Paluch, A. Drozd-Rzoska, J. Ziolo, P. Janik, K. Czupryński, Glassy and fluidlike behavior of the isotropic phase of n-cyanobiphenyls in broad-band dielectric relaxation studies, Eur. Phys. J. E 7 (2002) 387-392.

15. S.J. Rzoska, M. Paluch, S. Pawlus, A. Drozd-Rzoska, J. Ziolo, J. Jadzyn, Complex dielectric relaxation in supercooling and superpressing liquid-crystalline chiral isopentylcyanobiphenyl, Phys. Rev. E 68 (2003) 031795.

16. A. Drozd-Rzoska, S.J. Rzoska, S. Pawlus, J. Zioło, Complex dynamics of supercooling n-butylcyanobiphenyl (4CB), Phys. Rev. E 72 (2005) 031501.

17. A. Drozd-Rzoska, Glassy dynamics of liquid crystalline 4'-n-pentyl-4-cyanobiphenyl (5CB) in the isotropic and supercooled nematic phases, J. Chem. Phys. 130 (2009) 234910.

18. S.J. Rzoska and P.K. Mukherjee, M. Rutkowska, Does the characteristic value of the discontinuity of the isotropic–mesophase transition in n-cyanobiphenyls exist?, J. Phys.: Condens. Matter 24 (2012) 395101.

19. S.J. Rzoska, V. Mazur, A. Drozd-Rzoska (eds.), Metastable Systems under Pressure, Springer Verlag, Berlin, 2010.

20. S.J. Rzoska, A. Drozd-Rzoska, P.K. Mukherjee, D.O. Lopez, J.C. Martinez-Garcia, Distortion-sensitive insight into the pretransitional behavior of 4- n -octyloxy-4′-cyanobiphenyl (8OCB), J. Phys.: Condens. Matter 25, (2013) 245105.

21. A. Drozd-Rzoska, Quasi-tricritical' and glassy dielectric properties of a nematic liquid crystalline material, Crystals 10 (2020) 297.

22. J. Łoś, A. Drozd-Rzoska, S.J. Rzoska, S. Starzonek, K. Czupryński, Fluctuations-driven dielectric properties of liquid crystalline octyloxycyanobiphenyl and its nanocolloids, Soft Matter 18 (2022) 4502-4512.

23. S. Starzonek, A. Drozd-Rzoska, S.J. Rzoska, Pretransitional behavior and dynamics in liquid crystal' based nanocolloids, in M. M. Rahman, A. M. Asiri, Advances in Colloid Science, ch.12, pp. 265-274, InTech, Rijeka, 2016.

24. J.P.F. Lagerwall, Liquid Crystals with Nano and Microparticles, World Sci. Pub., Singapore, 2016.

25. I. Dierking, Nanomaterials in Liquid Crystals, MDPI, Basel, Switzerland, 2018.

26. S.J. Rzoska, S, Starzonek, A. Drozd-Rzoska, K. Czupryński, K. Chmiel, G. Gaura, A. Michulec, B. Szczypek, W. Walas, The impact of $BaTiO_3$ nanonoparticles on pretransitional effects in liquid crystalline dodecylcyanobiphenyl, Phys. Rev. E 93 (2016) 534.

27. S. Starzonek, S.J. Rzoska, and A. Drozd-Rzoska, K. Czupryński, S. Kralj, Impact of ferroelectric/superparaelectric nanoparticles on phase transitions and dynamics in nematic liquid crystal, Phys. Rev. E. 96, 022705 (2017).

28. A. Drozd-Rzoska, S. Starzonek, S. J. Rzoska, and S. Kralj, Nanoparticle-controlled glassy dynamics in nematogen-based nanocolloids, Phys. Rev. E 99 (2019) 052703.

29. D. Češnar, C. Kyrou, I. Lelidis, A.Drozd-Rzoska, S. Starzonek, S.J.Rzoska, Z. Kutnyak, S.Kralj, Impact of weak nanoparticle induced disorder on nematic ordering, Crystals 9 (2019) 171.

30. S.J Rzoska, S. Starzonek, J. Łoś, A. Drozd-Rzoska, S. Kralj, Dynamics and pretransitional effects in $C_{60}$ fullerene nanoparticles and liquid crystalline dodecylcyanobiphenyl (12CB) hybrid system, Nanomaterials 10 (2020) 2343.

31. J. Łoś, A. Drozd-Rzoska, S.J. Rzoska, S. Starzonek, K. Czupryński, Fluctuations-driven dielectric properties of liquid crystalline octyloxycyanobiphenyl and its nanocolloids, Soft Matter 18 (2022) 4502-4512.

32. J. Łoś, A. Drozd-Rzoska, S.J. Rzoska, K. Czupryński, The impact of ionic contribution to dielectric permittivity in 11CB liquid crystal and its colloids with $BaTiO_3$ nanoparticles, Eur. Phys. J. E 45 (2022) 74.

33. F. Schönhals and A. Kremer, Broadband Dielectric Spectroscopy, Springer, Berlin, Germanty, 2003.

34. H. Naito, Y. Yokoyama, S. Muraami, M. Imai, M. Okuda, A. Sugimura, Dielectric properties of nematic liquid crystals in low frequency regime, Mol. Cryst. Liq. Cryst. 262 (1995) 249-255.

35. A. Ghanadzadeh, M.S. Beevers, The low-frequency dielectric response of aligned supercooled nematic mixtures, J. Mol. Liq. 94 (2001) 97-112.

36. M. Mikułko, M.. Fraś, S. Marzec , J.Wróbel, M. D. Ossowska-Chruściel and J. Chruściel, Dielectric and Conductivity Anisotropy in liquid crystalline phases of strongly polar thioesters, Acta Phys. Polon. 113 (2008) 1155-1160.

37. N. Lebovka, T. Dadakova, L. Lysetsky, O. Melezhyk, G. Puchkovska, T. Gavrilko, J. Baran, M. Drozd, Phase transitions, intermolecular interactions and electrical conductivity behavior in carbon multiwalled nanotubes/nematic liquid crystal composites, J. Mol Struct. 887 (2008) 135-143.

38. S. Naemura, A. Sawada, Ionic conduction in nematic and smectic a liquid crystals, Mol. Cryst. Liq. Cryst. 400 (2010) 79-96.

39. P. Tadapatri, U.S. Heremath, C.V. Yelamaggad, S. Krishnamurthy, Permittivity, conductivity, elasticity, and viscosity measurements in the nematic phase of a bent-core liquid crystal, J. Phys. Chem. B, 114 (2010) 1745-175.

40. V.S. Chandel1, S. Manohar, J.P. Shukla, R. Manohar, A.K. Prajapati, Low frequency dielectric relaxation and optical behaviour of a nematic liquid crystal 4-methyl-n-(2'-hydroxy, 4'-n-hexadecyloxy azobenzene, Mat. Sci.-Poland, 30 (2012) 290-296.

41. F.-C. Lin, P.-C. Wu, B.-R. Jian, W. Lee, Dopant effect and cell-configuration-dependent dielectric properties of nematic liquid crystals, Adv. Condens. Matt. Phys. 2013 (2013).

42. S. Tomylko, O. Yaroshchuk, O. Kovalchuk, U. Maschke, R. Yamaguchi, Dielectric properties of nematic liquid crystal modified dielectric properties of nematic liquid crystal modified with diamond, Ukr. J. Phys. 57 (2012) 239-243.

43. S.K. Prasad, M.V. Kumar, T. Shilpa, C.V. Yelamagga, Enhancement of electrical conductivity, dielectric anisotropy and director relaxation frequency in composites of gold nanoparticle and a weakly polar nematic liquid crystal, RSC Adv. 4 4453-4462 (2014).

44. M. Urbanski, J.P.F. Lagerwall, Why organically functionalized nanoparticles increase the electrical conductivity of nematic liquid crystal dispersions, J. Mater. Chem. C 5 (2017) 8802-8809.

45. S. Patari, A. Nath, Tunable dielectric and conductivity properties of two 4-n alkoxy benzoic acid, Opto-Elect. Rev. 26 (2018) 35-43.

46. N. Dalir, S. Javadian, J. Kakemam, S.M. Sardpoor, Enhance the electrical conductivity and charge storage of nematic phase by doping 0D photoluminescent graphene was prepared with small organic molecule as a new array quantum dot liquid crystal displays, J. Mol. Liq. 276 (2019) 290-295.

47. M. Mrukiewicz, P. Perkowski, M. Urbańska, D.Węgłowska, W. Piecek, Electrical conductivity of ion-doped fluoro substituted liquid crystal compounds for application in the dynamic light scattering effect, J. Mol. Liq. 317 (2020) 113810.

48. G. Kaur, P. Kumar, A.K., Singh, D. Jayoti, P. Malik, Dielectric and electro-optic studies of a ferroelectric liquid crystal dispersed with different sizes of silica nanoparticles, Liquid Crystals 47 (2020) 2194-2208.

49. G. Kocakülah, M. Yıldırım, O. Köysal, İ. Ercan, Influence of UV Light intensity on dielectric behaviours of pure and dye-doped cholesteric liquid crystals. J. Mater. Sci. Mater. Electron. 31 (2020) 22385-22397.

50. S. Lalik, A. Deptuch, T. Jaworska-Gołąb, P. Fryń, D. Dardas, O. Stefańczyk, M. Urbańska, M. Marzec, Modification of AFLC physical properties by doping with $BaTiO_3$ particles. J. Phys. Chem. B 124 (2020) 60556-073.

51. D. Webb and Y. Garbovskiy, Overlooked ionic phenomena affecting the electrical conductivity of liquid crystals, Eng. Proc. 11 (2021) 1.

52. I.P. Studenyak, O.V. Kovalchuk, S.I. Peberezhets, M.M. Luchynets, A.I. Pogodin, M. Timko, P. Kopčaský, Electrical conductivity of composites based on 6CB liquid crystal and $(Cu_6PS_5I)_{0.5}(Cu_7PS_6)_{0.5}$ superionic nanoparticles, Mol. Cryst. Liq. Cryst. 718 (2021) 92-101.

53. H. Ayeb, T. Missaoui, A. Mouhli, F. Jomni, T. Soltani, Dielectric spectroscopy study on the impact of magnetic and nonmagnetic nanoparticles dispersion on ionic behavior in nematic liquid crystal, phase transit. 94 (2021) 37-46.

54. Y. Garbovskiy, Conventional and unconventional ionic phenomena in tunable soft materials made of liquid crystals and nanoparticles, Nano Express 2 (2021) 012004.

55. M.B. Salah, R. Nasri, A. N. Alharbi, T M. Althagafi, T. Soltani, Thermotropic liquid crystal doped with ferroelectric nanoparticles: electrical behavior and ion trapping phenomenon. J. Mol. Liq. 357 (2022) 119142.

56. N. Brouckaert, N. Podoliak, T. Orlova, D. Bankova, A.F. De Fazio, A G. Kanaras, O. Hovorka, G.D'Alessandro, and M. Kaczmarek, Nanoparticle-induced property changes in nematic liquid crystals, Nanomaterials 12 (2022) 341.

57. K. Takatoh, M. Sakamoto, R. Hasegawa, M. Koden, N. Itoh, M. Hasegawa (eds.), Alignment Technology and Applications of Liquid Crystal Devices, CRC Press, Routlege, 2019.

58. Q.S. Mei, and K.Lu, Melting and superheating of crystalline solids: from bulk to nanocrystals, Prog. Mater. Sci. 5 (2007) 1175-1262.

59. A.C. Lawson, Physics of the Lindemann rule, Phil. Mag. 89 (2009) 1757-1770.

60. Y. Yang, M. Asta, and B.B. Laird, Solid-liquid interfacial premelting, Phys. Rev. Lett. 110 (2013) 096102.

61. A. Samanta, M.E. Tuckerman, T.-Q. Yu, W. Ee, Microscopic mechanisms of equilibrium melting of a solid, Science 345 (2014) 729-732.

62. M. Pocheć, H. Niu, L. Ren, S. Bai, K. Orzechowski, Premelting phenomena in *n*-alcohols from nonanol to dodecanol, J. Phys. Chem. C 12 (2020) 21013–21017.

63. A. Drozd-Rzoska, S. Starzonek, S.J. Rzoska. J. Łoś, Z Kutnyak, Z.S. Kralj, Pretransitional Effects of the Isotropic Liquid–Plastic Crystal Transition, Molecules 26 (2021) 429.

64. A. Drozd-Rzoska, S.J. Rzoska, A. Szpakiewicz-Szatan, J. Łoś, K. Orzechowski, Pretransitional and premelting effects in menthol, Chem. Phys. Lett. 793 (2022) 139461.

65. A. Szpakiewicz-Szatan, S.J. Rzoska, A. Drozd-Rzoska, J. Kalabiński, Pretransitional behavior of electrooptic Kerr Effect in liquid thymol, Eur. Phys. J. E 45 (2022) 71.

66. A. Drzewicz, E. Juszyńska-Gałązka, M. Jasiurkowska-Delaporte, P. Kula, Insight into cold- and melt crystallization phenomena of a smectogenic liquid crystal, CrystEngComm 24 (2022) 3074-3087.

67. A. Selevou, G. Papamokos, M. Steinhart and G. Floudas, 8OCB and 8CB liquid crystals confined in nanoporousalumina: Effect of confinement on the structure and dynamics, J. Phys. Chem. B 121 (2017) 7382-7394.

68. Barium Titanate BaTiO$_3$ Nanoparticles / Nanopowder (BaTiO3, 99.9%, 50nm, Cubic).

https://www.us-nano.com/inc/sdetail/532 (accessed 2021-01-19).

69. S.J. Rzoska, J. Zioło, The nonlinear dielectric effect applied to study the pretransitional effects in the isotropic phase of the MBBA - benzene solutions, Acta Phys. Polon. A 78 (1990) 915 - 919.

70. S.J. Rzoska and J. Zioło, Critical properties in the vicinity of the critical consolute point for the 4-methoxybenzylidene-4'butylaniline - isooctane solution, Liquid Crystals 11 (1992) 9 -14.

71. S.W. Richardson, The so-called dielectric constant, Nature 117 (1926) 515.

72. N. Forbes, B. Mahon, Faraday, Maxwel and the Electromagnetic Field, How Two Men Revolutionized Physics, Prometheus book, NY, 2014.

73. A. Chełkowski, Dielectric Physics, PWN-Elsevier, 1990.

74. P. Bordewijk, On the relationship between the Kirkwood correlation factor and the dielectric permittivity, J. Chem. Phys. 73 (1980) 595-596.

75. A.Volmari, H. Weingaertner, Cross term and Kirkwood factors in dielectric relaxation in pure liquids, J. Mol. Liq. 98-99 (2002) 293-301.

76. S. Urban, J. Kędzierski, and R. Dąbrowski, Analysis of the dielectric anisotropy of typical nematics with the aid of the Maier-Meier equations, Zeit. Nurforsch. 55a (2000) 449-456.

77. J.C.R. Reis, T.P. Iglesias, Kirkwood correlation factors in liquid mixtures from an extended Onsager–Kirkwood–Fröhlich equation, Phys. Cem. Chem. Phys. 13 (2011) 10670-10680.

78. R. Mandle, S.J Cowling, I. Sage, M.E. Colclough, J.W Goodby, Relationship between molecular association and re-entrant phenomena in polar calamitic liquid crystals, J Phys Chem B 119 (2015) 3273-3280.

79. S.K. Kumar, S. Okudaira, M.K. Naoki, S. Yagihara, Broadband dielectric spectroscopy of a nematic liquid crystal in benzene, J. Chem. Phys. 129 (2008) 164509.

80. J. Jadżyn, J. Świergiel, I. Płowaś, R. Dąbrowski, U. Sokołowska, Dipolar aggregation and the static dielectric permittivity of some liquid crystalline materials, Ind. Eng. Chem. Res.52 (2013) 41094112.

81. S. Starzonek, S.J. Rzoska, A. Drozd-Rzoska, S. Pawlus, J.-C. Martinez-Garcia & L. Kiste*r*sky, Fractional Debye–Stokes–Einstein behaviour in an ultraviscous nanocolloid: glycerol and silver nanoparticles, Soft Matter 11 (2015) 5554-5562.

82. A. Drozd-Rzoska, Pressure-Related Universal previtreous behavior of the structural relaxation time and apparent fragility, Front. Mater. 6 (2019) 103.

83. A. Drozd-Rzoska, Activation volume in glass forming systems, Sci. Rep. 9 (2019)13787.